\documentclass[aip,preprint,amsmath]{revtex4-1}

\draft 

\usepackage{graphicx}
\usepackage{subcaption}
\usepackage{color}
\usepackage{MnSymbol}

\draft 

\definecolor{roig}{rgb}{0,0,0}

\begin{document}


\title{The Multigaussian method: a new approach to multichannel radiochromic film dosimetry} 

\author{I. M{\'e}ndez}
\email[]{nmendez@onko-i.si}
\author{A. Pol\v{s}ak}
\author{R. Hudej}
\author{B. Casar}
\affiliation{Department of Medical Physics, Institute of Oncology Ljubljana, Zalo\v{s}ka cesta 2, Ljubljana 1000, Slovenia}


\begin{abstract}

\textbf{Abstract:}

The main objective of multichannel radiochromic film dosimetry methods is to correct, or at least mitigate, spatial heterogeneities in the film-scanner response, especially variations in the active layer thickness. To this end, films can also be scanned prior to irradiation. In this study, the abilities of various single channel and multichannel methods to reduce spatial heterogeneities, with and without scanning before irradiation, were tested. A new approach to multichannel dosimetry, based on experimental findings, {\it i.e.}, the Multigaussian method, is introduced. The Multigaussian method assumes that the probability density function of the response vector formed by the pixel values of the different color channels, including irradiated and non-irradiated scans, follows a multivariate Gaussian distribution. The Multigaussian method provided more accurate doses than the other models under comparison, especially when incorporating the information of the film prior to irradiation.

\end{abstract}

\maketitle 

\section{Introduction}

Small fields, steep dose gradients or regions without electronic equilibrium are some of the many situations in which radiation doses can be measured using radiochromic films. Particular strengths of radiochromic films are weak energy dependence \cite{rink:2007, richter:2009, arjomandy:2010, lindsay:2010, massillon:2012, bekerat:2014}, near water-equivalence \cite{crijns:2102, aapm:55}, and high spatial resolution. On the other hand, the dosimetry system composed of radiochromic films and a flatbed scanner suffers from several sources of uncertainty, which can affect the film ({\it e.g.}, differences in the active layer thickness \cite{hartmann:2010}, evolution of film darkening with post-irradiation time \cite{andres:2010, chang:2014}, dependency on humidity and temperature \cite{rink:2008}, noncatalytic and ultraviolet-catalyzed polymerization \cite{girard:2012}, scratches), the scanner ({\it e.g.}, the warming-up of the lamp \cite{Paelinck:2007, ferreira:2009}, inter-scan variations \cite{lewis:2015, mendez:2016}, noise \cite{bouchard:2009, vanHoof:2012}, dust), or the interaction between film and scanner ({\it e.g.}, the lateral artifact \cite{Schoenfeld:2014, vanbattum:2015, schoenfeld:2016b}, the dependency on the orientation of the film on the scanner bed \cite{butson:2009}, the dependency on film-to-light source distance \cite{lewis:2015, palmer:2015}, Newton rings \cite{dreindl:2014}, the cross talk effect \cite{vanbattum:2015}). Film handling and scanning protocols \cite{aapm:55, bouchard:2009}, corrections \cite{lewis:2012, mendez:2013, lewis:2015, lewis:2015b, mendez:2016, ruiz:2017} and multichannel dosimetry methods \cite{AMicke:2011, mayer:2012, mendez:2014, perez:2014} have been developed to reduce uncertainties and deliver precise and accurate doses \cite{palmer:2013, vera:2018}.

Multichannel radiochromic film dosimetry methods combine the information provided by all three color channels (R, G and B) of the scanner. In order to do so, the perturbation of the film-scanner response caused by different sources of uncertainty must be (explicitly or implicitly) modeled. Current multichannel methods assume that changes in the response for the different channels are positively correlated. Moreover, perturbations should be small. For example, even though the lateral artifact produces positively correlated perturbations ({\it i.e.}, pixel values decrease with the distance from the center of the scan for all three channels, given a constant dose), these perturbations increase with the dose and the distance from the center, eventually becoming excessive for multichannel models \cite{schoenfeld:2016b}. The primary targets of multichannel film dosimetry are the spatial heterogeneities of the response and, in particular, differences in the active layer thickness. These differences can also be addressed by incorporating the information of the film prior to irradiation employing net optical density (NOD) as the film-scanner response. 

The purpose of this study was to select the most accurate method for mitigating spatial heterogeneities in the response. The effectiveness of NOD and various multichannel models was compared with a novel approach to multichannel radiochromic film dosimetry which we have called the Multigaussian method.

\section{Methods and materials}

\subsection{Measurements}

Commonly, multichannel models are evaluated using the gamma index: a sample of clinical plans are calculated with the treatment planning system (TPS) and posteriorly irradiated, dose distributions are computed with each multichannel model and the multichannel model that produces dose distributions which are most similar to the TPS doses according to the gamma index is deemed to be the most accurate model \cite{mendez:2014, perez:2014, mendez:2015}. Film corrections are also examined in a similar fashion \cite{menegotti:2008, palmer:2015}. This procedure has several drawbacks \cite{Clasie:2012}: apart from the accuracy of the multichannel model, the gamma index depends on the shape of the dose distribution in the individual plans, the noise in the distributions, the tolerance criteria, the interpolation method implemented for the gamma index computation, the choice of reference and evaluation dose distributions, the accuracy of the TPS, etc. 

In this work, the measurement protocol was designed with the aim of minimizing the influence of any source of uncertainty other than the spatial heterogeneity of the response. Two lots of Gafchromic EBT3 films (lots 06061401 and 04071601) and one lot of EBT-XD films (lot 12101501) were employed. They were identified as lots A, B and C, respectively. The dosimetry system was completed with an Epson Expression 10000XL flatbed scanner (Seiko Epson Corporation, Nagano, Japan) and the scanning software Epson Scan v3.49a. Five films per lot were scanned prior to and 24 h following irradiation. Irradiation was delivered with a 6 MV beam from a Novalis Tx accelerator (Varian, Palo Alto, CA, USA). Films were positioned at source-axis distance (SAD) on top of the IBA MatriXX detector (IBA Dosimetry GmbH, Germany) inside the IBA MULTICube phantom. Doses were simultaneously measured with the MatriXX detector. Field dimensions were 20$\times$20 ${\rm cm^2}$ for lot A and 25$\times$25 ${\rm cm^2}$ for lots B and C. Films were irradiated with 1, 2, 4, 8 and 16 Gy for lots A and B, and 2, 4, 8, 16 and 32 Gy for lot C. In this setup, 1 Gy corresponded to 109 MU.

Before use, the scanner was warmed up for at least 30 min. Films were centered on the scanner bed with a transparent frame. In order to correct inter-scan variations, a 20.32$\times$4.50 ${\rm cm^2}$ unexposed fragment was kept in a constant position and scanned together with the films. Films were scanned in portrait orientation ({\it i.e.}, the scanner lamp was parallel to the short side of the film). The distance between film and lamp was kept constant by placing a 3 mm thick glass sheet on top of the film \cite{lewis:2015, palmer:2015}. Scans were acquired in transmission mode with a resolution of 50 dpi. Processing tools were disabled. Ten scans were taken for each film, the first five were discarded and the resulting image was the median of the remaining five scans. Images were saved as 48-bit RGB format (16 bit per channel) TIFF files. 

\subsection{Data set}
From the measurements, each pixel of each film was associated with a position $(x, y)$, three (R, G and B) pixel values (PVs) prior to irradiation, three PVs after irradiation, and the dose measured with MatriXX at that position. MatriXX dose arrays, which have a resolution of 7.62 mm/px, were bicubically interpolated to the resolution of the films. In order to avoid large dose gradients, which are associated with higher uncertainties, points with a measured dose lower than 95\% of the nominal dose were excluded from the set. Data analysis was implemented in the R environment \cite{R:software}.

Inter-scan and lateral corrections were applied on irradiated and non-irradiated PVs. Inter-scan variations were corrected using the unexposed fragment, following the column correction method \cite{mendez:2016}. Lateral corrections were applied according to the model
\begin{equation}
v(x) = \alpha_1 (x-x_c) + \alpha_2 (x-x_c)^{2} + \hat{v}(x) (1 + \beta_1 (x-x_c) + \beta_2 (x-x_c)^{2})
\end{equation}
where $\hat{v}$ represents the PV before the lateral correction, $x$ is the coordinate on the axis parallel to the lamp, $x_c$ is the position of the center of the scanner, and $v$ are corrected PVs \cite{mendez:2013, lewis:2015b}. The images of unexposed films can facilitate the determination of the parameters in this equation. If we symbolize with $\hat{v_0}(x)$ and $v_0(x)$ the PV as a function of $x$ of unexposed films before and after applying lateral corrections, we can rewrite the model as
\begin{equation}
\label{lat-corr-zero}
v(x) = v_0(x) + (\hat{v}(x) - \hat{v_0}(x)) (1 + \beta_1 (x-x_c) + \beta_2 (x-x_c)^{2})
\end{equation}
Apart from being trivially obtained, another advantage of using $\hat{v_0}(x)$ and $v_0(x)$ is that they are robust against variations in the active layer, since they can be calculated from the average (or median) of several films. For each color channel, $\hat{v_0}(x)$ and $v_0(x)$ in equation~\ref{lat-corr-zero} were fitted from the median of the film scans prior to irradiation, and the $\beta$ parameters were derived from the irradiated films. The corrected PVs ($v$) are necessary for computing the $\beta$ parameters. Corrected PVs were derived from measured doses. The relationship between PVs and measured doses ({\it i.e.}, the calibration) was calculated using regions of interest (ROIs) with dimensions 3$\times$3 ${\rm cm^2}$ centered on the fields. Sensitometric and inverse sensitometric curves for all three color channels were modeled with natural cubic splines. 

Therefore, apart from the positions, measured irradiated PVs, measured non-irradiated PVs, and measured doses, the data set included PVs after applying inter-scan corrections and PVs after applying inter-scan and lateral corrections.

\subsection{The Multigaussian method}
Perturbations of the response in radiochromic film dosimetry can be expressed as
\begin{equation}
D(r) = D_{k}(z_{k}(r) + \Delta_k(r)) 
\end{equation}
where $D_k$ denotes the calibration function for color channel $k$, $z(r)$ is the film-scanner response at point $r$, and $\Delta$ is the perturbation.

\begin{figure}
\begin{minipage}[b]{0.9\linewidth}
\centering
\includegraphics[width=\linewidth]{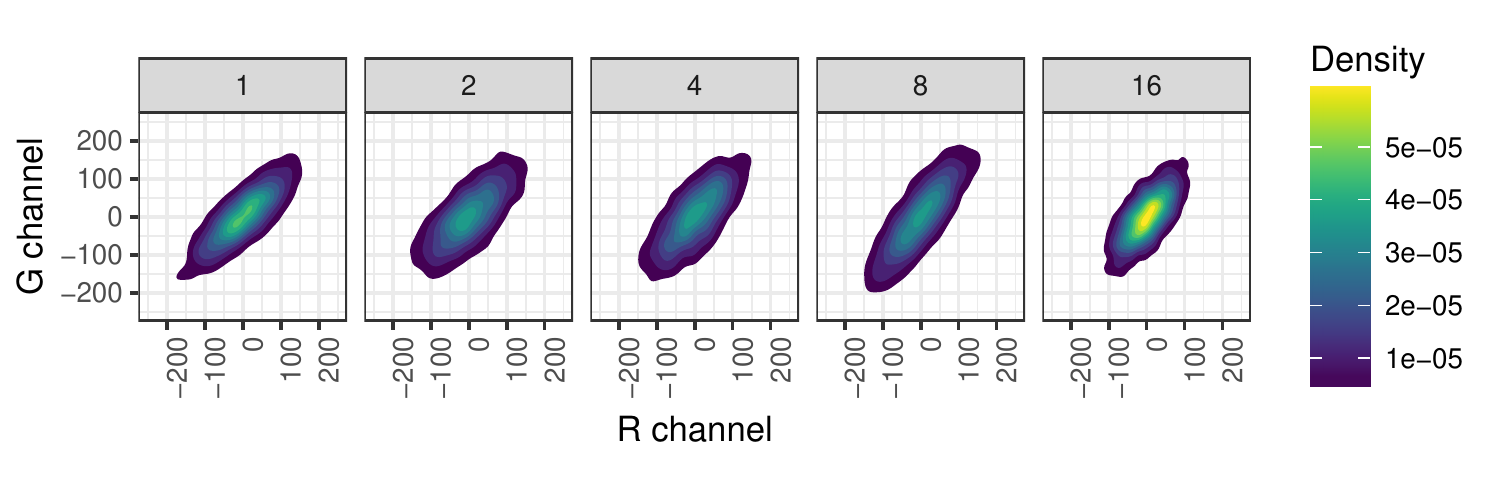}\\
\end{minipage}
\begin{minipage}[b]{0.9\linewidth}
\centering
\includegraphics[width=\linewidth]{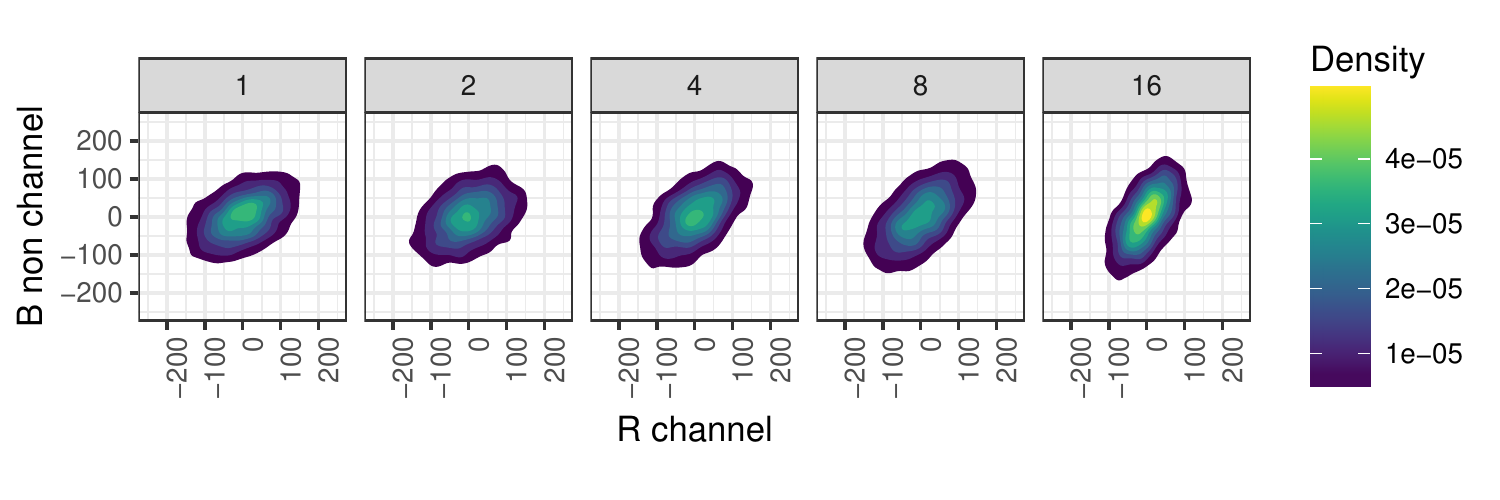}\\
\end{minipage}
\begin{minipage}[b]{0.9\linewidth}
\centering
\includegraphics[width=\linewidth]{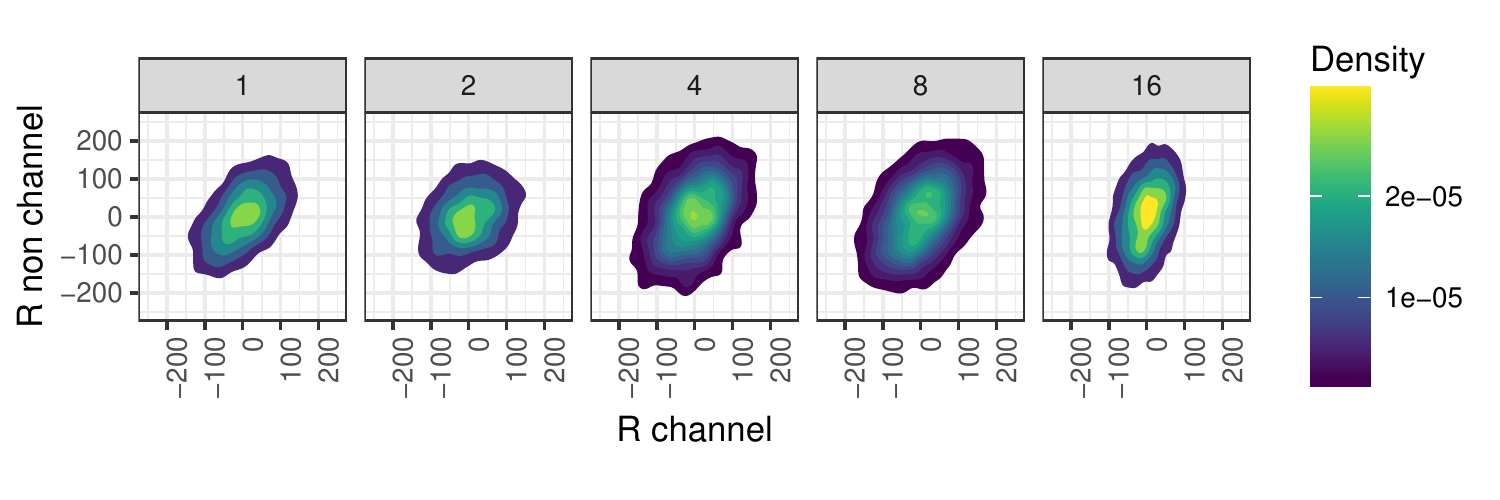}\\
\end{minipage}
\begin{minipage}[b]{0.9\linewidth}
\centering
\includegraphics[width=\linewidth]{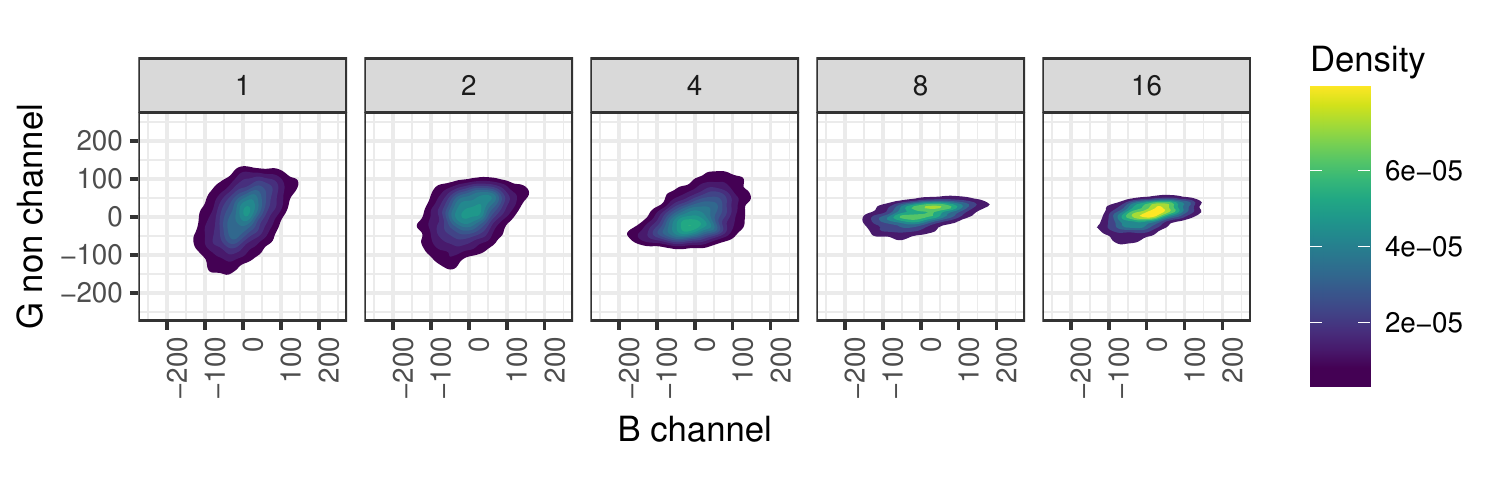}\\
\end{minipage}
\caption{\label{fig:perturbations} Joint PDF of perturbations in PVs for different combinations of color channels (X and Y axes) and different doses (in Gy, represented in the columns).}
\end{figure}

Figure~\ref{fig:perturbations} shows the plots of the joint probabilities of perturbations ({\it i.e.}, $P(\Delta_i, \Delta_j)$) for different combinations of color channels and different doses. The perturbations are calculated as the differences between the mean PV and the PV of each point in the ROIs used for the calibration. Measured PVs were analyzed prior to irradiation (channels R non, G non and B non) and after irradiation (channels R, G and B). For illustration purposes, only four combinations of channels using the films from lot A are shown. However, every possible combination of channels for every lot and film was examined. It can be observed that the perturbations are small (in general, lower than 2\% of the response) and positively correlated, and that the joint probabilities might be approximated with bivariate Gaussian distributions. This behavior points to a novel approach to radiochromic film dosimetry: the Multigaussian method. 

The Multigaussian method considers that, given a dose $D$, the probability of the response vector $\mathbf{z}$ ({\it i.e.}, the vector with the responses $z_k$ for each channel) obeys a multivariate Gaussian distribution 
\begin{equation}
\label{eq:multigaussian}
P(\mathbf{z} \mid D) \sim \mathcal{N}_k(\boldsymbol{\mu}(D), \mathbf{\Sigma}(D)) 
\end{equation}
Here, $k$ is the number of different channels ({\it i.e.}, irradiated channels and optionally non-irradiated channels), $\boldsymbol{\mu}$ is the vector of expected values of the response and $\mathbf{\Sigma}$ is the covariance matrix  
\begin{equation}
\Sigma_{ij} = cov[z_i, z_j] = E[(z_i - \mu_i)(z_j - \mu_j)]
\end{equation}

During the calibration, $\boldsymbol{\mu}(D)$ and $\mathbf{\Sigma}(D)$ can be measured for a set of doses. For the remaining doses, they should be interpolated. In this study, we interpolated $\mu_k(D)$ and $\Sigma_{ij}(D)$ with natural cubic splines.  

Following a Bayesian approach, the probability of each dose $D$ given a response vector $\mathbf{z}$ can be computed as 
\begin{equation}
P(D \mid \mathbf{z}) = \frac{P(\mathbf{z} \mid D) P(D) }{P(\mathbf{z})} \propto P(\mathbf{z} \mid D) 
\end{equation}
where the prior probabilities of $D$ are considered to be equiprobable. 

Therefore, given a response $\mathbf{z}$, the dose (or more precisely the probability density function of the dose) can be obtained by
\begin{equation}
P(D \mid \mathbf{z}) \propto \frac{\exp \left(-\frac{1}{2} (\mathbf{z}-\boldsymbol{\mu}(D))^{T} \mathbf{\Sigma}(D)^{-1} (\mathbf{z}-\boldsymbol{\mu}(D))\right)}{\sqrt{(2\pi)^k|\mathbf{\Sigma}(D)|}}
\end{equation}

\subsection{Model selection}
The Multigaussian method was compared with other multichannel and single channel film dosimetry models. The relevance of scanning before irradiation was also evaluated. The Multigaussian method can integrate the information from the film prior to irradiation as additional dimensions of the response vector ($z_k$) in terms of PVs. In this work, the Multigaussian method was applied with and without non-irradiated channels, and the other models were calculated both with PVs and NODs as response. The additional use of optical density (OD) was discarded since there is only a change of coordinates between PV and OD. NOD was computed as
\begin{equation}
z := \log_{10} \frac{v_{\rm{non}}}{v_{\rm{irr}}}
\end{equation}
where $v_{\rm{irr}}$ indicates the PVs of the irradiated film, and $v_{\rm{non}}$ indicates the PVs of the film prior to irradiation. 

Many models were compared, but for the sake of clarity, only the most successful of them are analyzed here. The best results with single channel models were obtained using the red (R) channel. While, among the multichannel models, the most successful were the Mayer model \cite{mayer:2012} and what we called the Normal Factor model. 

The Mayer model considers that the perturbation is additive and approximately equal for all channels:
\begin{equation}
D(r) = D_{k}(z_{k}(r) + \Delta(r) + \delta_{k}(r)) 
\end{equation}
where $\delta_{k}$ symbolizes the error term. The perturbation is presumed to be small. Thus, a first-order Taylor expansion of the dose in terms of the channel independent perturbation yields
\begin{equation}
D(r) = D_{k}(z_{k}(r)) + \dot{D}_{k}(z_{k}(r))\Delta(r) + \epsilon_{k}(r) 
\end{equation}
where $\dot{D}_{k}$ is the first derivative of $D_{k}$ with respect to $z_{k}$ and $\epsilon_{k}$ is an error term. Different assumptions about the probability density functions (PDFs) of $\Delta$ or $\epsilon_{k}$ lead to different estimations of the dose \cite{mendez:2014}. The Mayer model considers that the PDF of $\Delta$ is a uniform distribution and that all $\epsilon_{k}$ follow the same normal distribution.The most likely value of the dose ($d$) in the Mayer model is calculated as
\begin{equation}
d = \frac{(\sum_{k=1}^{n} \dot{D}_{k}) (\sum_{k=1}^{n} D_{k} \dot{D}_{k}) - (\sum_{k=1}^{n} \dot{D}_{k}^{2}) (\sum_{k=1}^{n} D_{k})}{(\sum_{k=1}^{n} \dot{D}_{k})^{2} - n (\sum_{k=1}^{n} \dot{D}_{k}^{2})}
\end{equation}  

The Normal Factor model assumes a multiplicative perturbation which is approximately equal for all channels: 
\begin{equation}
D(r) = D_{k}(\Delta(r)z_{k}(r) + \delta_{k}(r)) 
\end{equation}
Again, different PDFs for $\Delta$ or $\delta_{k}$ give rise to different dose distributions. In the Normal Factor model, all $\delta_{k}$ follow the same normal distribution ({\it i.e.}, $\delta_{k} \sim \mathcal{N}(0, \sigma_{\delta}^2)$), but the PDF of $\Delta$ is also normal ({\it i.e.}, $\Delta \sim \mathcal{N}(1, \sigma_{\Delta}^2)$). In this case, the most likely value of the dose is 
\begin{equation}
\label{eq-normalfactor}
d = \underset{D}{\arg\max} \, P(D) = \underset{D}{\arg\min} \left( \frac{(\Delta - 1)^2}{\sigma_{\Delta}^2} + \frac{\sum_k \delta_{k}^2}{\sigma_{\delta}^2} \right) = \underset{D}{\arg\min} \left( (\Delta - 1)^2 + \frac{1}{\gamma}\sum_k \delta_{k}^2 \right)
\end{equation} 
where $\gamma$ is the ratio $\sigma_{\delta}^2/\sigma_{\Delta}^2$. The best results for the Normal Factor model were obtained with $\gamma = 5\times10^8$, which is roughly the square of the mean PV, meaning that both $\Delta$ and $\delta_k$ contributed to the perturbation with similar weight. In order to calculate the dose $d$, it can be deduced from equation~\ref{eq-normalfactor} that, given $D$, 
\begin{equation}
\Delta = \frac{\gamma + \sum_k z_k \mu_k}{\gamma + \sum_k z_k^2}
\end{equation} 
and
\begin{equation}
\delta_{k} = \mu_k  - \Delta z_k
\end{equation} 
where $\mu_k$ is obtained from the calibration $D=D_k(\mu_k)$. 

\section{Results and discussion}

\begin{figure}
\begin{minipage}[b]{0.9\linewidth}
\centering
\includegraphics[width=\linewidth]{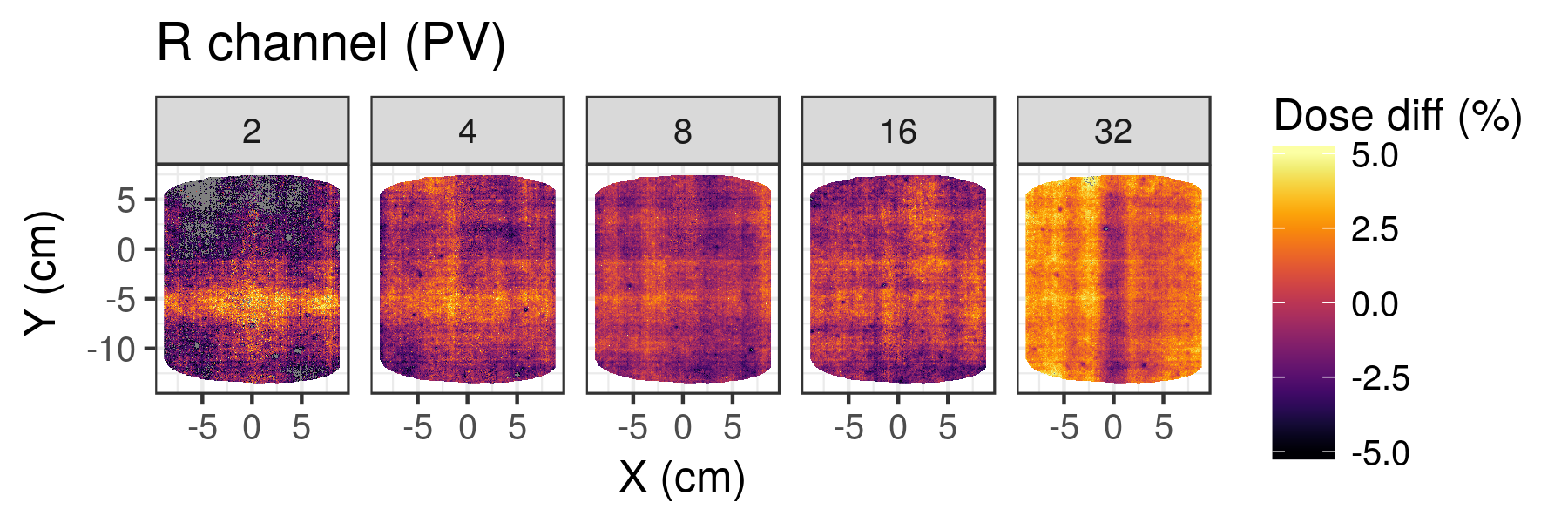}\\
\end{minipage}
\begin{minipage}[b]{0.9\linewidth}
\centering
\includegraphics[width=\linewidth]{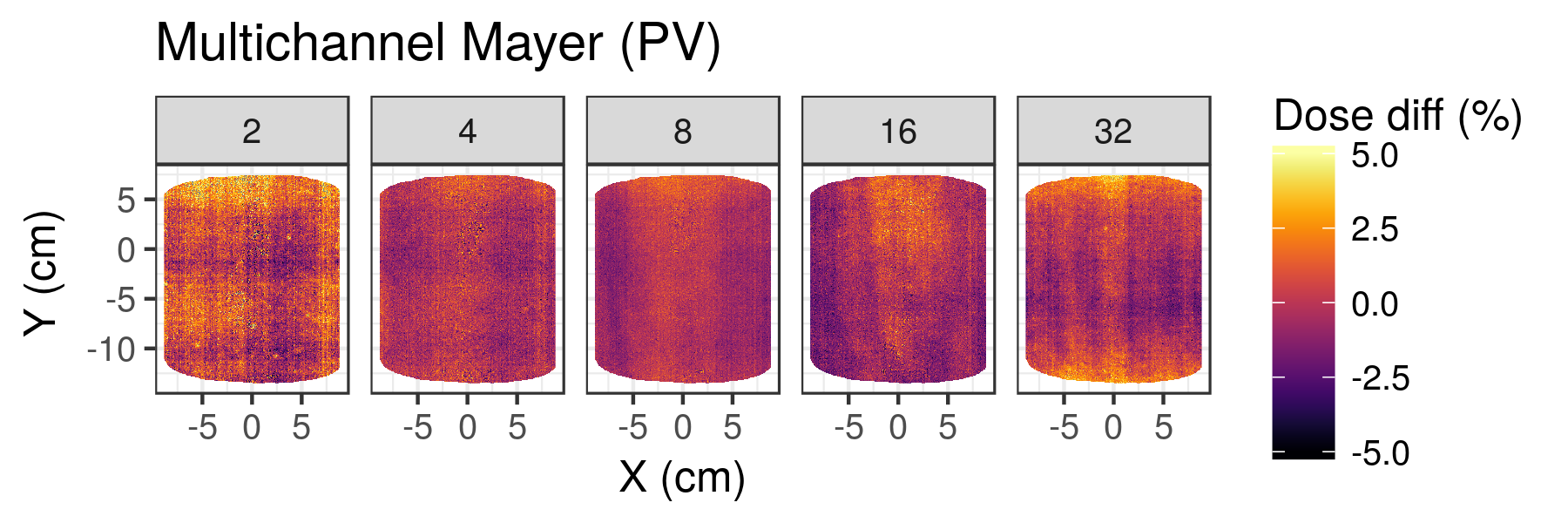}\\
\end{minipage}
\begin{minipage}[b]{0.9\linewidth}
\centering
\includegraphics[width=\linewidth]{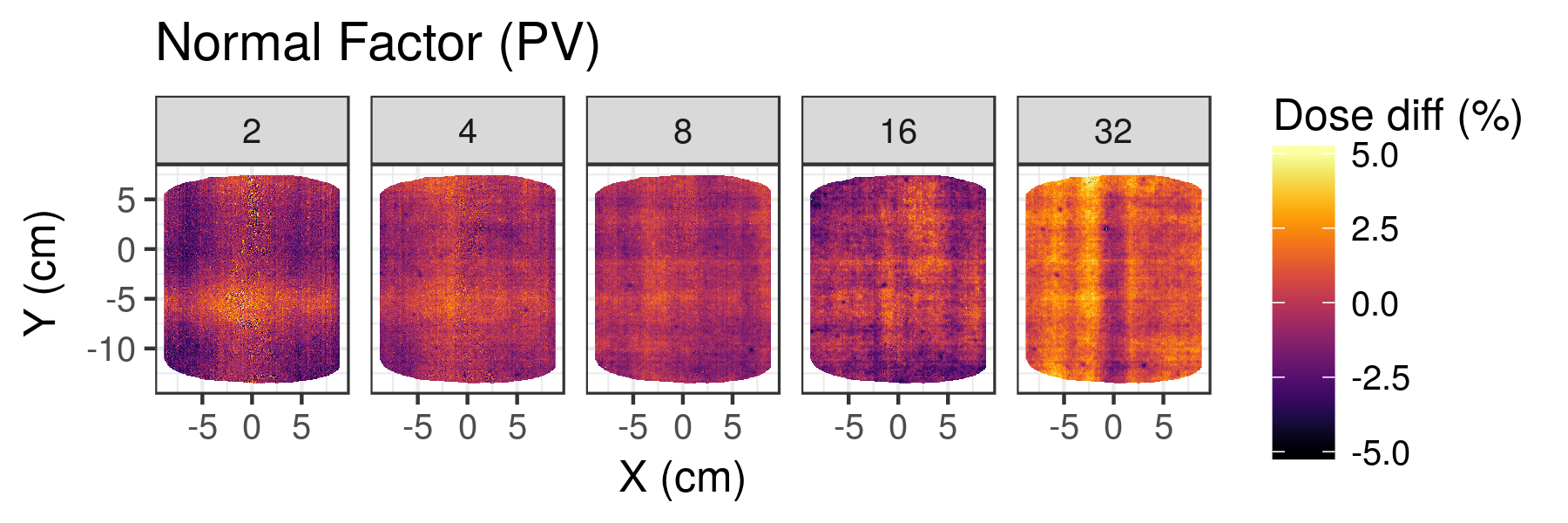}\\
\end{minipage}
\begin{minipage}[b]{0.9\linewidth}
\centering
\includegraphics[width=\linewidth]{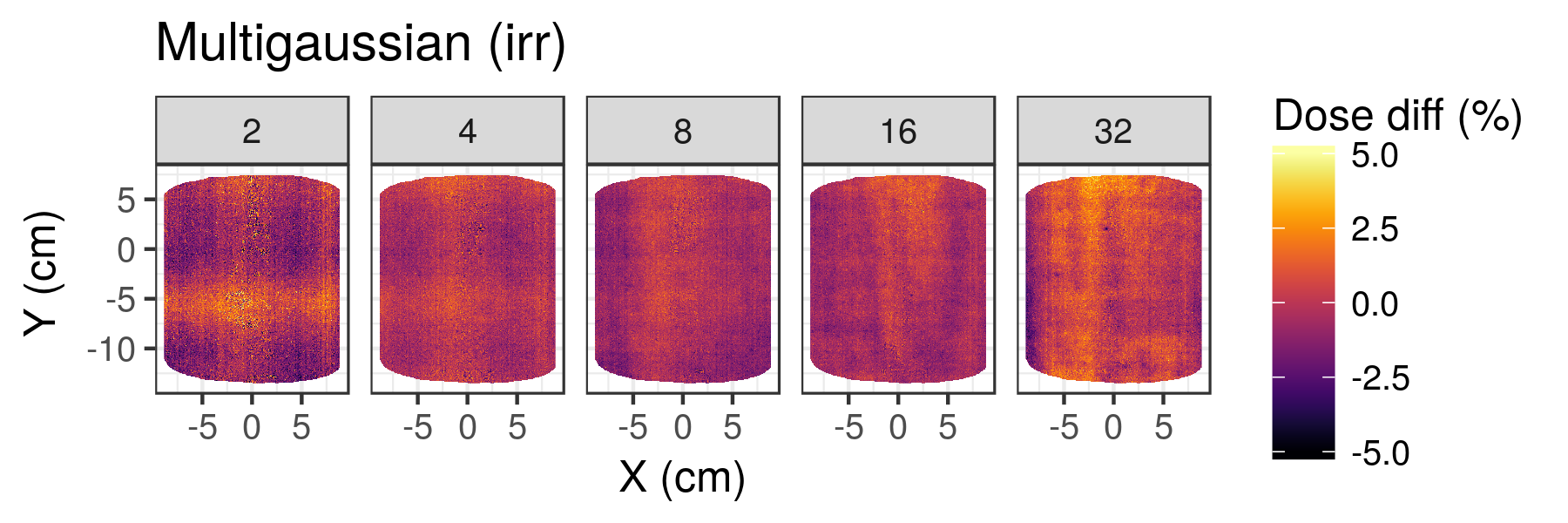}\\
\end{minipage}
\caption{\label{fig:models-pv} Dose differences between measured doses and doses calculated with each film dosimetry model. Irradiated PVs after applying inter-scan and lateral corrections were employed as response. Films from lot C are showed. Each column plots one film designated by its nominal dose (in Gy).}
\end{figure}

\begin{figure}
\begin{minipage}[b]{0.9\linewidth}
\centering
\includegraphics[width=\linewidth]{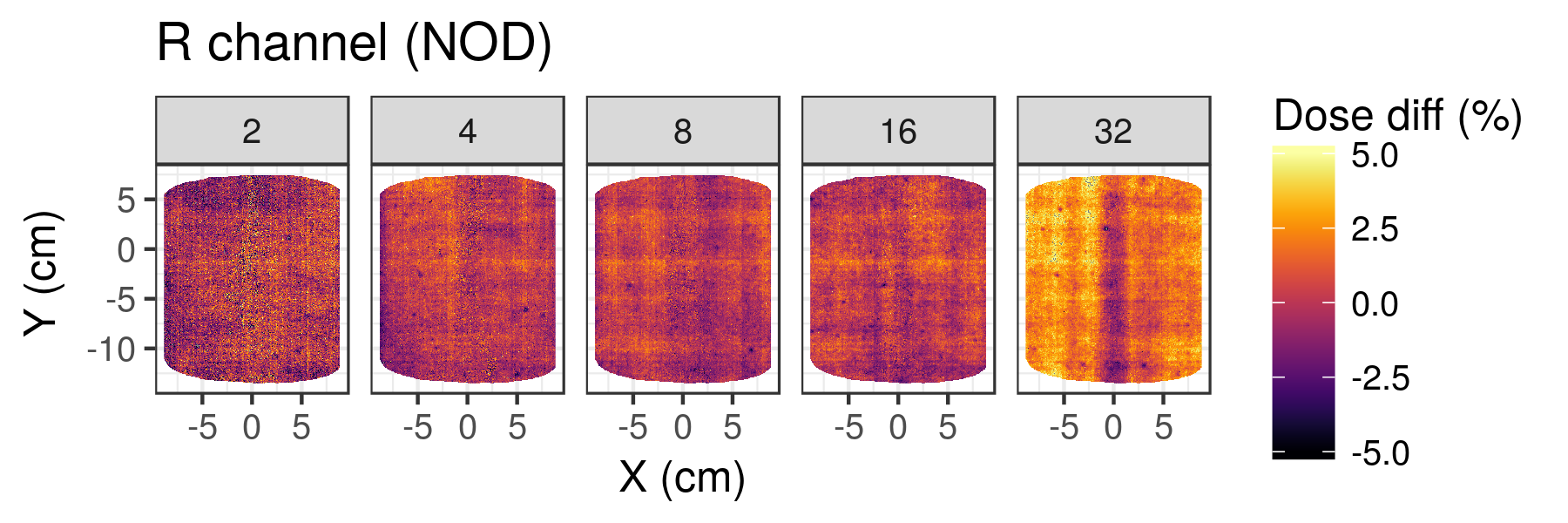}\\
\end{minipage}
\begin{minipage}[b]{0.9\linewidth}
\centering
\includegraphics[width=\linewidth]{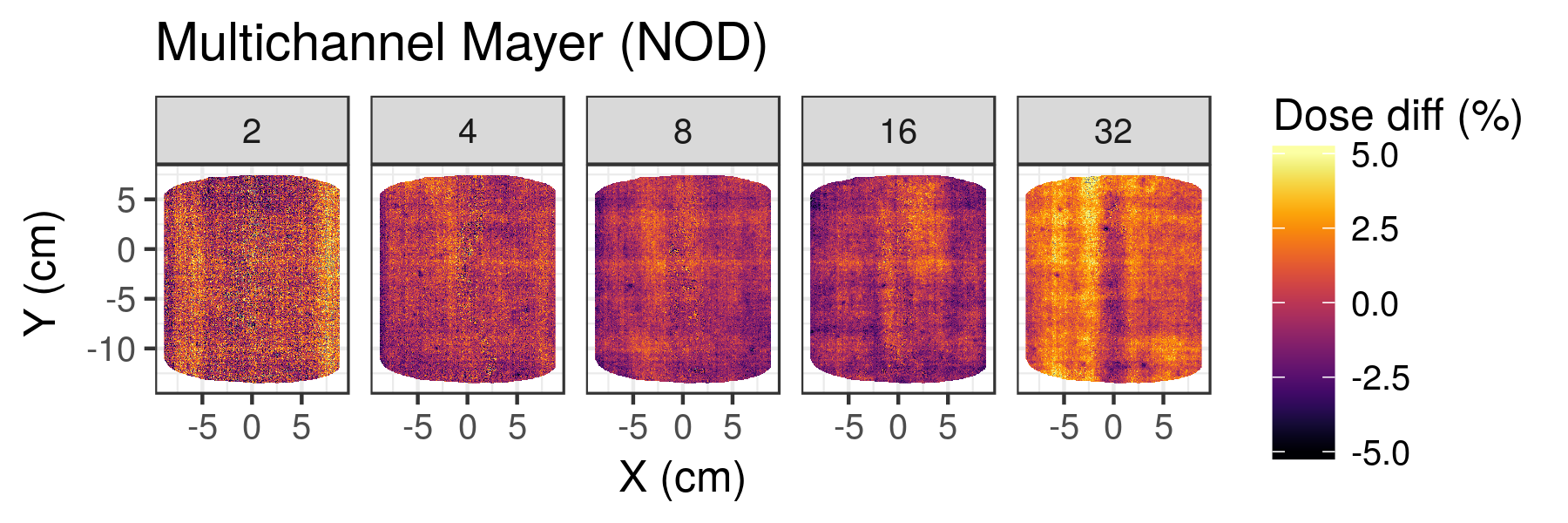}\\
\end{minipage}
\begin{minipage}[b]{0.9\linewidth}
\centering
\includegraphics[width=\linewidth]{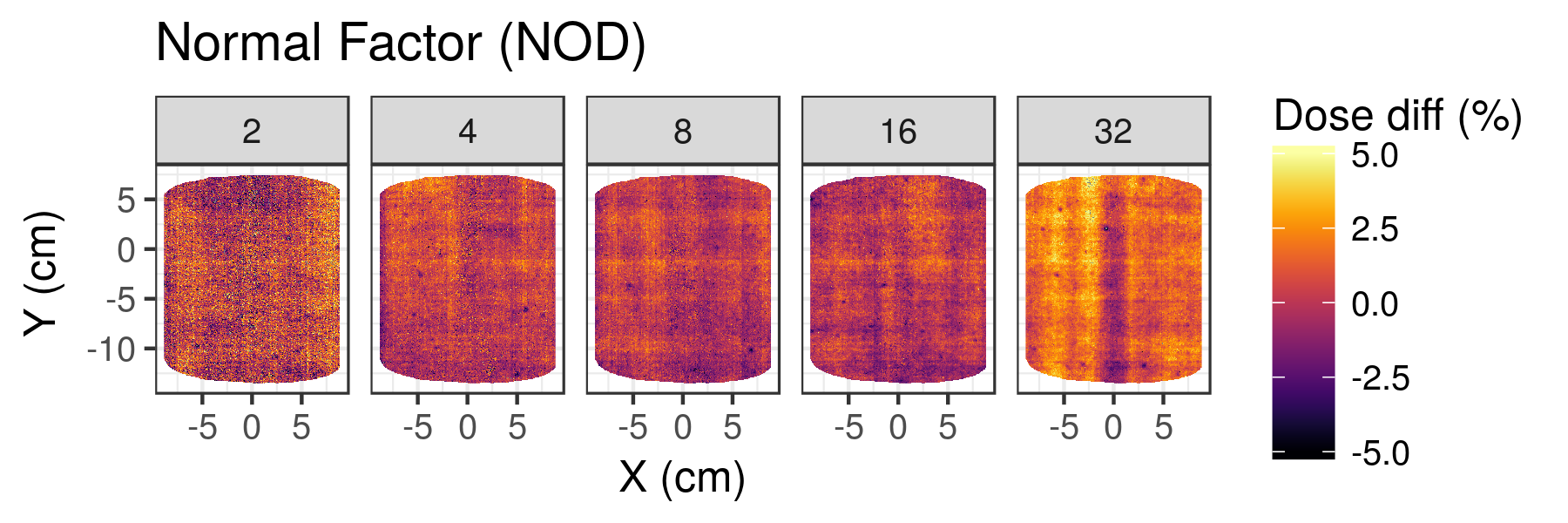}\\
\end{minipage}
\begin{minipage}[b]{0.9\linewidth}
\centering
\includegraphics[width=\linewidth]{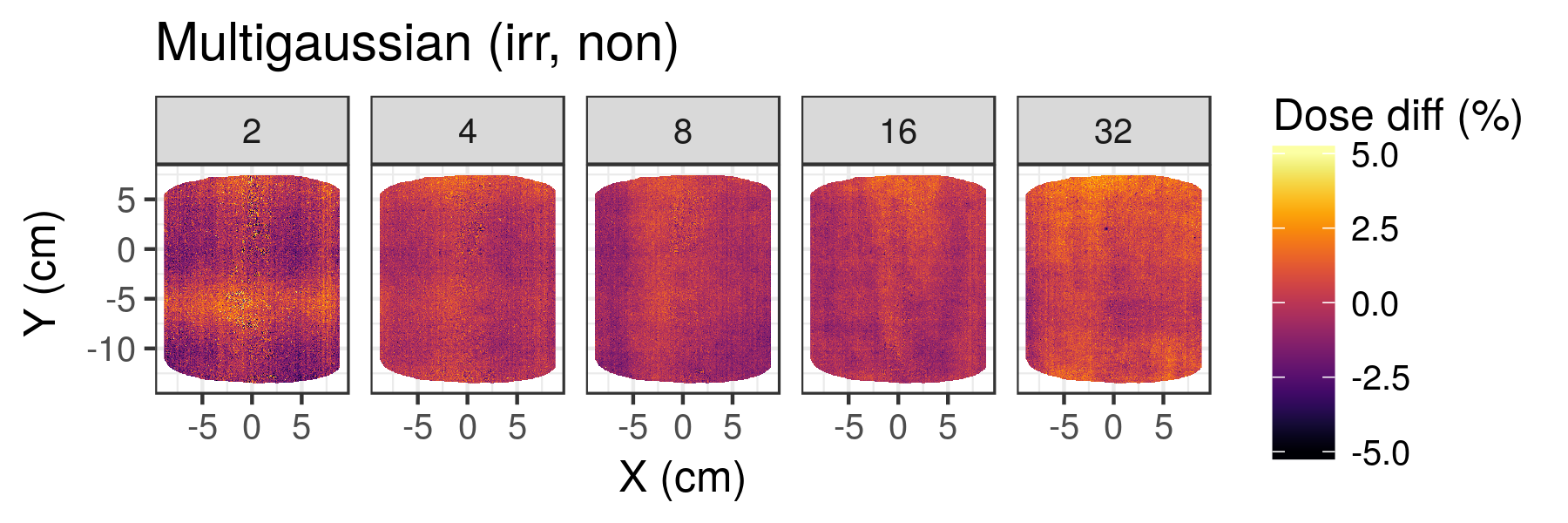}\\
\end{minipage}
\caption{\label{fig:models-nod} Dose differences between measured doses and doses calculated with each film dosimetry model. After applying inter-scan and lateral corrections, NODs or irradiated and non-irradited PVs for the Multigaussian model were employed as response. Films from lot C are showed. Each column plots one film designated by its nominal dose (in Gy).}
\end{figure}

Figure~\ref{fig:models-pv} and figure~\ref{fig:models-nod} show the relative dose differences between measured doses with MatriXX and doses calculated with each dosimetry model. In figure~\ref{fig:models-pv}, irradiated PVs were employed as the response. In figure~\ref{fig:models-nod}, the Multigaussian model used both irradiated and non-irradiated PVs and the rest of the models employed NODs. In both figures, inter-scan and lateral corrections were applied. Although only films from lot C are shown, all three lots produced similar outcomes. Table~\ref{tab:dose_diffs} condenses this information into the standard deviation of the dose difference between measured and calculated doses for each lot. 

It can be observed that the Multigaussian model delivered more accurate doses. However, no method corrects spatial heterogeneities completely. Low frequency heterogeneities, presumably due to active layer variations, are still visible \cite{gonzalez:2017}. Whether using only the scans of irradiated films, or including the scans prior to irradiation, and for each lot under study, the Multigaussian model gave lower dose differences than the other three models. Scanning prior to and after irradiation improved the results of the Multigaussian and R channel models, but this was not true for all lots when employing the Mayer or Normal Factor models. Therefore, in this study, the most accurate film dosimetry model for mitigating spatial heterogeneities was the Multigaussian method scanning prior to and after irradiation. 

\begin{table}
\caption{\label{tab:dose_diffs} Standard deviations (k = 1) of the relative dose differences (\%) between the dose measured with MatriXX and the dose measured with films.}
\begin{ruledtabular}
\begin{tabular}{lccc}
Film dosimetry method & Lot A & Lot B & Lot C  \\
\hline 
R channel (PV)          	& 1.7 & 2.3 & 2.0 \\
Multichannel Mayer (PV)		& 1.6 & 2.4 & 1.4 \\
Factor Normal (PV)      	& 1.6 & 1.7 & 1.3 \\
Multigaussian (irr)     	& 1.4 & 1.4 & 1.1 \\
\hline
R channel (NOD)             & 1.2 & 1.7 & 1.6 \\
Multichannel Mayer (NOD)    & 1.6 & 2.0 & 1.8 \\
Factor Normal (NOD)        	& 1.3 & 1.8 & 1.6 \\
Multigaussian (irr, non) 	& 1.1 & 1.2 & 1.0 \\
\end{tabular}
\end{ruledtabular}
\end{table}

Table~\ref{tab:dose_diffs_lat} shows how uncertainties increase when lateral corrections are not applied. In this case, however, it should be noted that giving a unique value for the standard deviation of the dose differences is not informative enough, since the differences increase with the dose and the distance from the center. In spite of this, the results support the recommendation of applying lateral corrections even with multichannel film dosimetry methods.
 
\begin{table}
\caption{\label{tab:dose_diffs_lat} Standard deviations (k = 1) of the relative dose differences (\%) between the dose measured with MatriXX and the dose measured with films when lateral corrections are not applied.}
\begin{ruledtabular}
\begin{tabular}{lccc}
Film dosimetry method & Lot A & Lot B & Lot C  \\
\hline 
R channel (PV)           & 4.0 & 4.4 & 3.2 \\
Multichannel Mayer (PV)  & 2.0 & 2.1 & 1.9 \\
Factor Normal (PV)       & 2.5 & 2.9 & 2.1 \\
Multigaussian (irr)      & 2.0 & 2.6 & 1.5 \\
\hline
R channel (NOD)          & 3.5 & 3.5 & 2.5 \\
Multichannel Mayer (NOD) & 3.6 & 4.2 & 2.9 \\
Factor Normal (NOD)      & 1.9 & 2.4 & 2.1 \\
Multigaussian (irr, non) & 1.5 & 2.0 & 1.3 \\
\end{tabular}
\end{ruledtabular}
\end{table}

\section{Conclusions}

The purpose of this work was to select the method which best reduces spatial heterogeneities in radiochromic film dosimetry. The design of the experiment aimed to minimize any other sources of uncertainty. Single channel and multichannel methods, including a novel multichannel approach to film dosimetry called the Multigaussian method, were compared. The convenience of scanning both prior to and after irradiation was examined also. 

The Multigaussian method provided more accurate doses than the other models. The best results were obtained with the Multigaussian method integrating the information from the film before irradiation. This information is not integrated as net optical density but as additional elements in the response vector, which is composed of the pixel values of the different color channels including irradiated and non-irradiated scans. The assumption of the Multigaussian method is that the probability density function of the response vector follows a multivariate Gaussian distribution. This assumption is based on empirical findings.

Although the Multigaussian method obtained the best agreement between doses measured with film and MatriXX, spatial heterogeneities, presumably caused by variations in the active layer thickness, are still present. Therefore, new and more accurate multichannel radiochromic film dosimetry methods are necessary. This research demonstrates a new way of tackling the problem.

\begin{acknowledgments}
I. M{\'e}ndez is co-founder of Radiochromic.com. 
\end{acknowledgments}

\providecommand{\noopsort}[1]{}\providecommand{\singleletter}[1]{#1}%

\end{document}